\documentclass[12pt]{iopart}

\usepackage[]{units}
\usepackage{acronym}
\usepackage[T1]{fontenc}
\usepackage{lettrine} 
\usepackage{gensymb} 
\usepackage{float} 
\usepackage{nomencl}
\usepackage{caption}
\usepackage{subcaption}
\usepackage{ragged2e}
\usepackage{multirow}
\newcommand{\ssymbol}[1]{^{\@fnsymbol{#1}}}
\makeatother
\usepackage{graphicx}
\usepackage[colorinlistoftodos]{todonotes}
\usepackage{hyperref} 
\usepackage{booktabs,caption}
\usepackage[flushleft]{threeparttable}
\hypersetup{colorlinks = true,linkcolor = blue,anchorcolor =red,citecolor = blue,filecolor = red,urlcolor = blue}
\graphicspath{ {./img/} }

\acrodef{IR}[IR]{Infra-Red}
\acrodef{INFN}[INFN]{Istituto Nazionale di Fisica Nucleare}
\acrodef{PCTO}[PCTO]{Percorsi per le Competenze Trasversali e per l'Orientamento}
\acrodef{PLS}[PLS]{Piano Lauree Scientifiche}
\begin{document}

\title[Teaching Physics by Arduino during COVID-19 Pandemic: The Free Falling Body Experiment]{Teaching Physics by Arduino during COVID-19 Pandemic: The Free Falling Body Experiment}

\author{Fausto Casaburo}

\address{Sapienza Università di Roma, Dipartimento di Fisica}
\address{Istituto Nazionale di Fisica Nucleare (INFN) Sezione Roma}
\ead{fausto.casaburo@uniroma1.it-fausto.casaburo@roma1.infn.it}
\vspace{10pt}
\begin{indented}
\item[]June 2021
\end{indented}

\begin{abstract}
One of the difficulties related to the COVID-19 pandemic is the shifting from face-to-face to distance teaching. Both schools and universities had suddenly to organize on-line lectures. To perform laboratory practice even in this period, easily accessible materials, smartphones physics apps, on- line tools and devices can be used. In this paper a method to measure the gravitational acceleration studying the free falling body using Arduino board is presented.

\end{abstract}

\vspace{2pc}

%
%
%


\tableofcontents

\section*{Introduction}\label{Introduction}
\addcontentsline{toc}{section}{Introduction}
\lettrine[nindent=0em,lines=3]{T}he COVID-19 pandemic had a huge impact on teaching worldwide. Despite that, the educational
system responded shifting from face-to-face to distance learning \cite{PhysRevPhysEducRes.17.010117}. 

For preserving laboratory courses, despite the restrictions introduced by the COVID-19 pandemic, many educational institutions overcame this problem by enabling the students to perform physics experiments at home \cite{doi:10.1119/5.0020515}.

For example, in Italy, the Lab2Go project \cite{Lab2Gowebpage,Lab2GoWiki,ORGANTINI2017PRO} proposed on-line  seminars  aimed  at showing experiments that can be made at home using easily accessible materials and exploiting resources as the Arduino board \cite{Arduino}.

Arduino is an open source platform made of electronic boards, sensors and expansion boards.  Its usage also allows to acquire additional competences, as for example coding and programming \cite{Organtini_2018, Organtini_fisica_arduino}. The original Arduino board can be bought for as low as about few tens of euro; moreover, there are many cheapest clones. Both the board and the sensors can be easily bought on-line or shops of electronic components \cite{Organtini_fisica_arduino}. There are also many kits including the board and common sensors available for just 50-60 euro. Thanks to the low cost, the Arduino board and related component can be bought directly by students or by schools/universities to be provided to students.

In this paper an Arduino-based physics experiment regarding of the free falling body will be presented. It consists in bringing down a ball and measuring the time of flight by Arduino to estimate the gravitational acceleration value. 
 
 The experiment can be proposed both to high school and university students.

\section{Theory}\label{Theory}
 When an object is only subjected to the gravitational force, it is called free falling body. The motion of a free falling body moving along the $y$-axis is described by the equation:
 
\begin{equation}
y\left(t\right)=y_{0}+v_{0}t-\frac{1}{2}gt^{2}
\label{eq:ffb}
\end{equation}

where $g=\unitfrac[9.80665]{m}{s^{2}}$ \cite{Valore_g_doi:10.1063/1.555817} is the average gravitational acceleration. In this experiment it hasn't been used a vacuum tube but, for small heights, the resistance of the air can be assumed negligible.

\section{Experimental setup}\label{Setup}
\begin{figure}[H]
\centering
\includegraphics[width=4in]{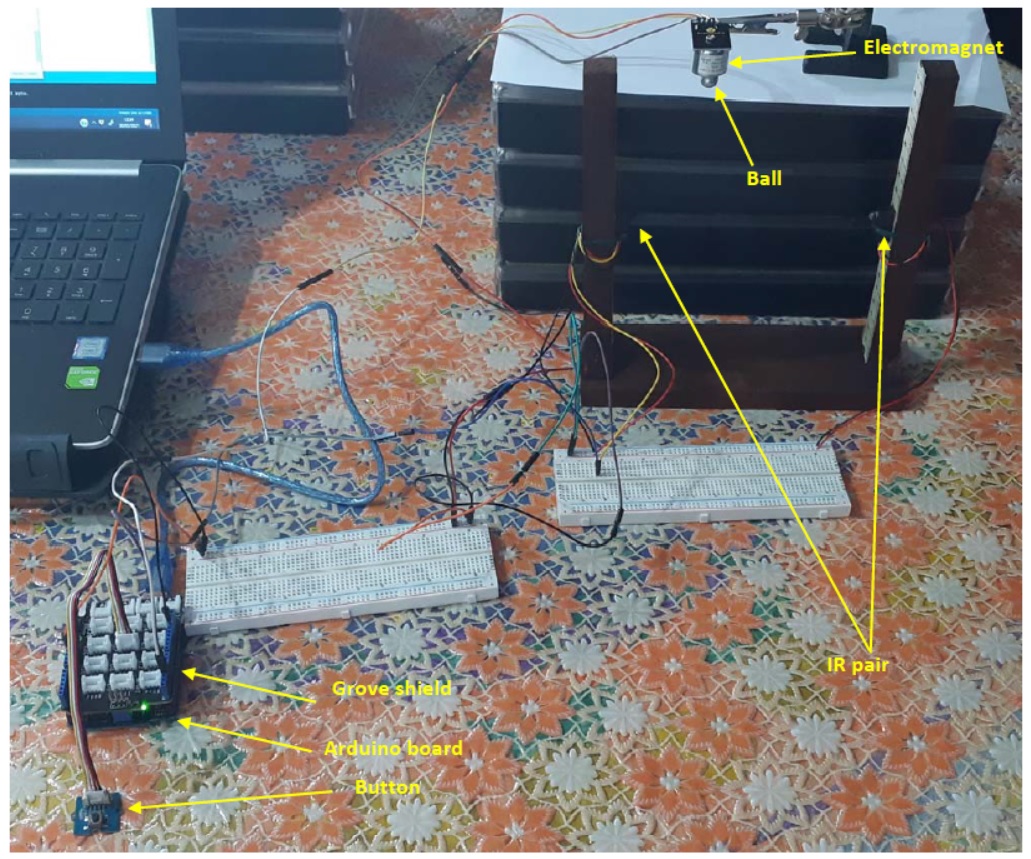}\DeclareGraphicsExtensions.
\caption{Experimental setup. }
\label{fig:exp_setup}
\end{figure}

The experimental setup (Fig. \ref{fig:exp_setup}) consists of an Arduino UNO R3 board, a button connected to Arduino by a Grove Shield (not mandatory, but usefull to simplify connections), an electromagnet, an \ac{IR} trasmitter-receiver pair sensors, a soldering support, boxes, metallic marble, meterstick, breadboards, Dupont cables, USB cable, a support having U-shape and one computer. {\color{black}{ The common button (i.e. the one not requiring the shield), the Duponts cables and the \ac{IR} pair sensor could be found in a typical Arduino kit. On the contrary, the electromagnet (and, if you want to use it, the shield and the specif button) must be bought separately.}}

\begin{figure}[H]
\centering
\includegraphics[width=3in]{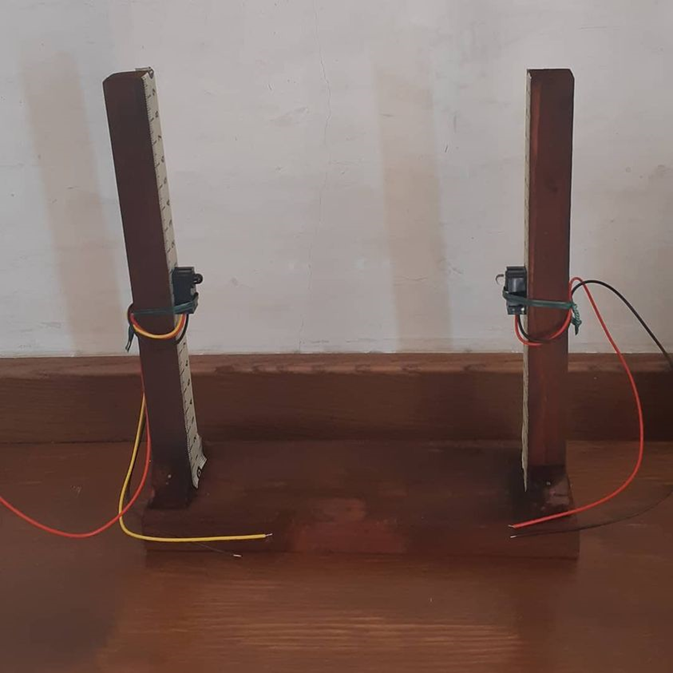}\DeclareGraphicsExtensions.
\caption{U-shape support for \ac{IR} pair sensors.}
\label{fig:u_support}
\end{figure}

The electromagnet is connected to Arduino by Dupont cables and it is supported in parallel to the table by the soldering support that, in turn, is positioned on the boxes. Their number is variable  in order to change the initial height of the marble with respect to \ac{IR}  sensors that are supported by the hand-made U-shape support (Fig. \ref{fig:u_support}) positioned underside of the electromagnet. {\color{black}{Notice that, the maximum current that could be supplied by Arduino is $\unit[40]{mA}$; therefore, if your electromagnet needs a higher current, it must be powered by an external power supply, not directly by the board.}}
Falling down, the marble moves through the \ac{IR} sensors and the time of flight can be measured. 

The electromagnet, the button and the \ac{IR} sensors are connected to Arduino board by its digital pins. The overall setup is connected to the computer for data acquisition by the USB cable.

\section{Experimental procedure}\label{Procedure}

The sketch allows the user to control the electromagnet by the button: pushing it, the digital pin is setted to HIGH state and it's possible to attach the marble; pushing it again the pin is switched to LOW state therefore the marble falls down due to gravitational force. {\color{black} The \ac{IR} light is constantly emitted from the transmitter of the the \ac{IR} pair sensors and, if there aren't obstacle, it will hit the \ac{IR} receiver and it is read by the Arduino board as a HIGH state of the pin whose the sensor is connected. When the marble moves through the beam light, this one is interrupted, therefore the pin state of the sensor will be switched to LOW state, allowing to measure the time between the fall and passage through the beam. Notice that we are assuming a perfect passage of the marble through the beam, but this is just approximation and a potential error. Anyawy, in order to obtain best possible passage, it would be helpful using a transparent plastic tube to guide the initial flight of the ball.}

\begin{figure}[H]
\centering
\includegraphics[width=5in]{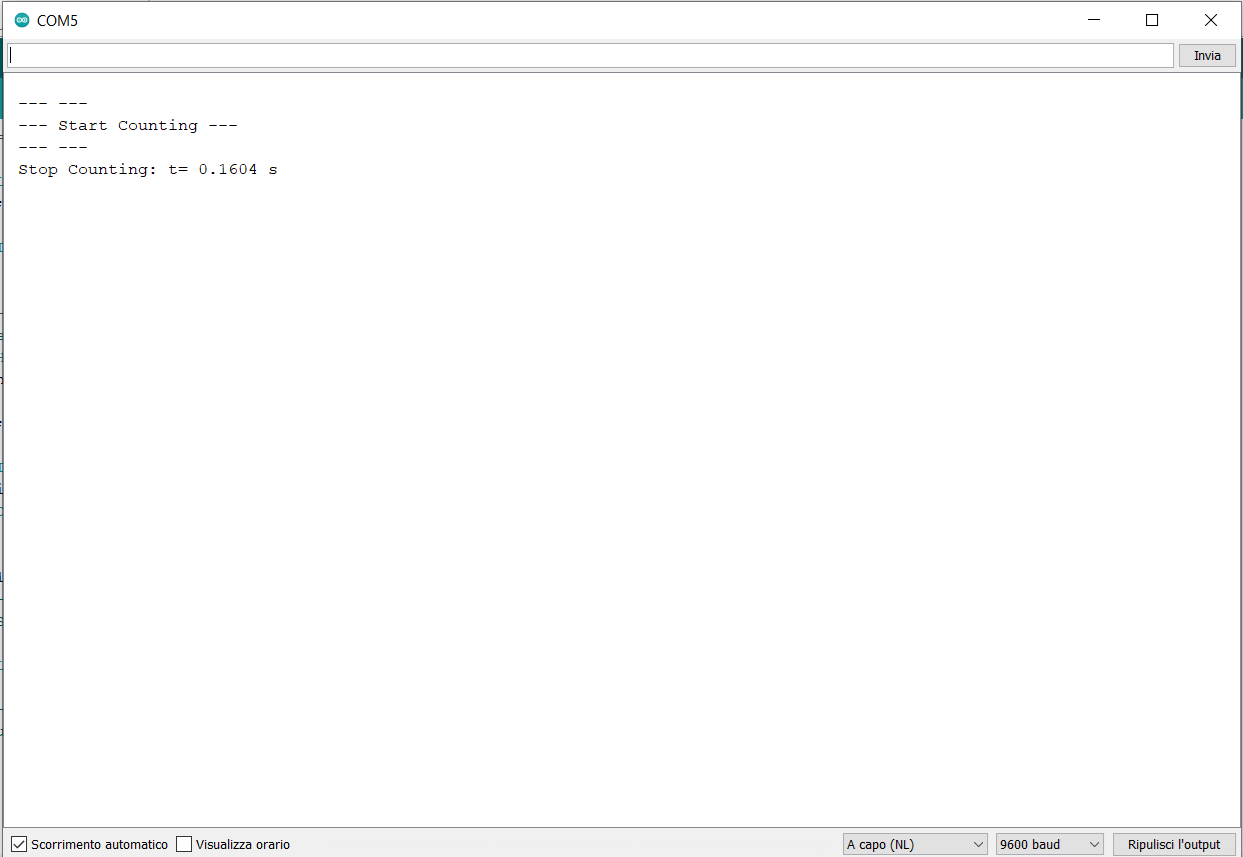}\DeclareGraphicsExtensions.
\caption{Example of time of flight measurement by Arduino.}
\label{fig:falling_time}
\end{figure}

Defining $t_i$ and $t_f$ respectively the times when the marble falls down and when it moves through the \ac{IR} sensors, the time of flight is given by:

\begin{equation}
\triangle t=t_{f}-t_{i}
\label{eq:fall_time}
\end{equation}

and it is measured by Arduino and printed on the terminal (Fig. \ref{fig:falling_time}). {\color{black}{Notice that, formula \ref{eq:fall_time}  assumes that the marble falls down exactly at the moment when the button is pushed. Yet, in reality, the demagnetization of the electromagnet is affected by a bit delay due to its hysteresis; also the responce of the \ac{IR} pair would be affected by a little delay. Therefore, for a more accurate measure, it would be usefull to look for it in the datasheet or to ask these delay times to the producer in order to take account of them.}}

The equation of motion is given by:

\begin{equation}
h\left(\triangle t\right)=\frac{1}{2}g\triangle t^{2}
\label{eq:equation_motion}
\end{equation}

therefore, measuring by the meterstick the distance between the initial position of the marble and \ac{IR} sensors, the $g$ value is given by: 

 \begin{equation}
g=\frac{2h\left(\triangle t\right)}{\triangle t^{2}}
\label{eq:g_formua}
\end{equation}

The times of flight for several heights have been measured and analyzed. 
In particular, data of the heights in fuction of $\triangle t^{2}$ have been drawn and interpulated by the linear function:

\begin{equation}
h\left(x\right)=mx
\label{eq:linear_function}
\end{equation}

where $x=\triangle t^{2}$ and $m=\frac{1}{2}g$. Lastly, the gravitational acceleration is given by:

\begin{equation}
g=2m
\label{eq:linear_function}
\end{equation}

\section{Results}\label{Results}

\begin{figure}[H]
\centering
\includegraphics[width=5in]{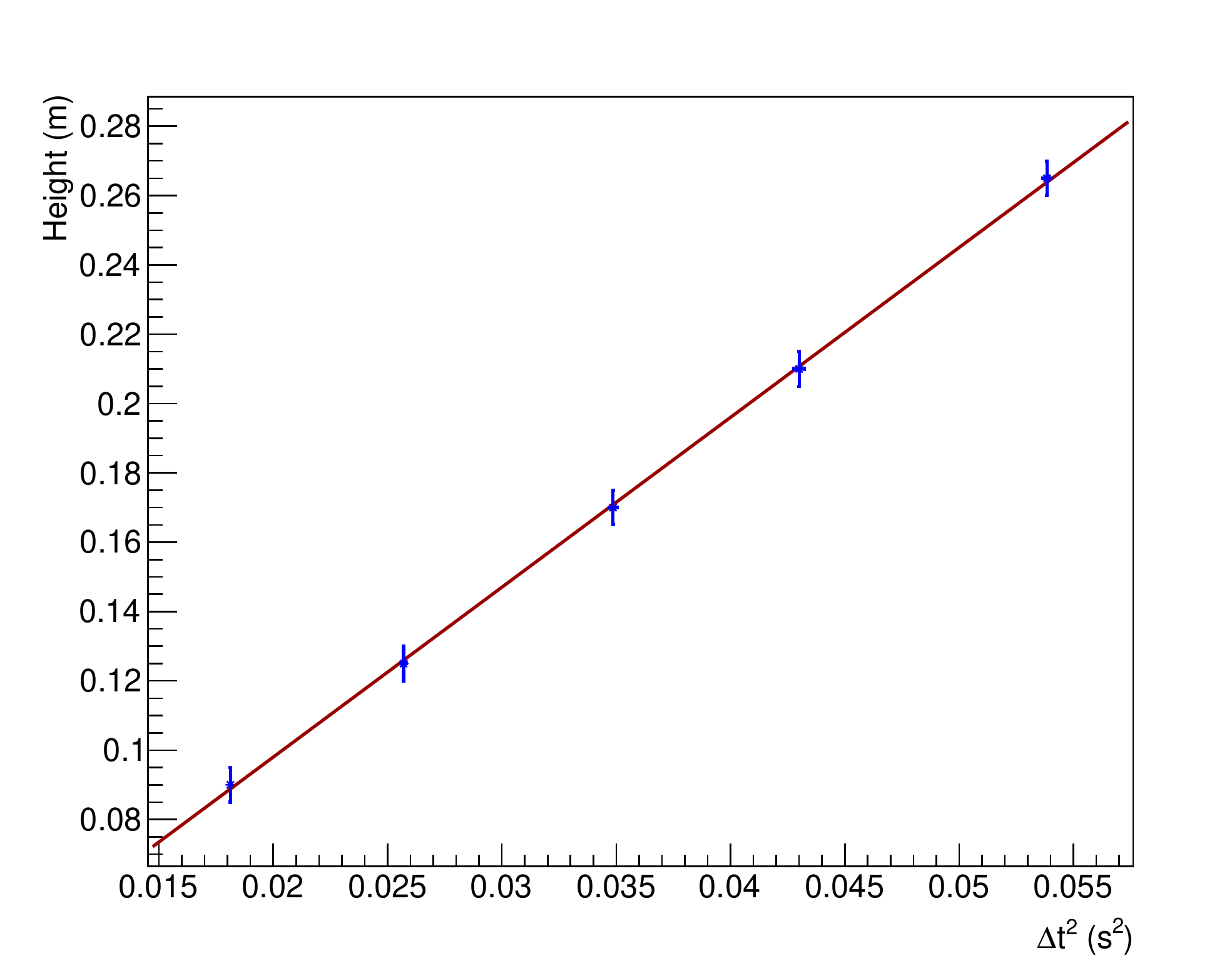}\DeclareGraphicsExtensions.
\caption{Fit.}
\label{fig:fit}
\end{figure}

The graph of the height in function of  $\triangle t^{2}$ is shown in Fig. \ref{fig:fit}. 

The gradient got by fit and the obtained $g$ value are given in Tab. \ref{tab:results}

\begin{table}[H]
\centering
 \caption{ Fit result gradient and obtained $g$ value.}
 \label{tab:results} 
  \begin{threeparttable}
     \begin{tabular}{ll}
        \toprule
        \(\unit[m]{\left(\unitfrac{m}{s^{2}}\right)}\)  & \(\unit[g]{\left(\unitfrac{m}{s^{2}}\right)}\) \\
        \midrule
	         $4.88\pm0.35$   &  $9.8\pm0.7$\\
        \bottomrule
     \end{tabular}
  \end{threeparttable}
\end{table}

The obtained $g$ value is compatible with the one in literature within $1\sigma$.

\section*{Conclusions}
\addcontentsline{toc}{section}{Conclusions}
One of the several problems due to COVID-19 pandemic, in educational framework, is the impossibility to access to laboratories. A key issue is to organize laboratory
activities made at home using easily accessible materials and exploiting resources as smartphones physics apps, on-line tools and devices. In this context, the Arduino board has great importance because it allows to perform physics experiments even at home.

In this paper it has been shown a technique to measure the gravitational acceleration using  Arduino. Beyond the numerical result, the article goal is to encourage teachers to propose the experiment to their students in order to carry on the laboratory practice even in this pandemic period. 

Lastly, thanks to its low-cost, the usage of Arduino for physics experiments can also be useful in school laboratories not adequately equipped (obsolete or non-functioning instrumentation, poor assortment, lack in maintenance, missing catalog) even when the COVID-19 emergency is over.

\section*{Acknowledgements}
\addcontentsline{toc}{section}{Acknowledgements}
The author acknowledges the Lab2Go- Fisica collaboration in particular, professors Pia Astone, Giulia De Bonis, Riccardo Faccini, Giovanni Organtini and Francesco Piacentini.
\section*{References}
\bibliographystyle{ieeetr}
\bibliography{mybiblio}

\acrodef{LED}[LED]{Light Emitting Diode}

\end{document}